\documentclass{elsart}
\usepackage{epsfig}
\usepackage{graphicx}

\begin{document}
\runauthor{Ramos, Fernando Manuel}
\begin{frontmatter}
\title{Atmospheric turbulence within and
above an Amazon forest}
\author[INPE]{Fernando Manuel Ramos\corauthref{Someone}}
\author[UNIVAP]{Maur\'{\i}cio Jos\'e Alves Bolzan}
\author[INPE,MUSEU]{Leonardo Deane de Abreu S\'a}
\author[INPE]{Reinaldo Roberto Rosa}
\address[INPE]{Instituto Nacional de Pesquisas Espaciais, INPE \\
               S\~ao Jos\'e dos Campos - SP, Brazil}
\address[UNIVAP]{Universidade do Vale do Para\'{\i}ba, UNIVAP \\
                 S\~ao Jos\'e dos Campos - SP, Brazil}
\address[MUSEU]{Museu Paraense Em\'{\i}lio Goeldi (Campus de Pesquisa) \\
                Coordena\c{c}\~ao de Ci\^encias da Terra e Ecologia (CCTE), Escrit\'orio do INPE \\
                Bel\'em - PA, Brazil}
\corauth[Someone]{Corresponding Author: fernando@lac.inpe.br}

\begin{abstract}

In this paper, we discuss the impact of a rain forest canopy on the statistical characteristics of atmospheric turbulence. 
This issue is of particular interest for understanding on how the Amazon terrestrial biosphere interact with the atmosphere. 
For this, we used a probability density function model of velocity and temperature differences based on Tsallis' non-extensive 
thermostatistics. We compared theoretical results with experimental data measured in a 66 m micrometeorological tower, during the 
wet-season campaign of the Large Scale Biosphere-Atmosphere Experiment in Amazonia (LBA). Particularly, we investigated how the 
value of the entropic parameter is affected when one moves into the canopy, or when one passes from day/unstable to 
night/stable conditions. We show that this new approach provides interesting insights on turbulence in a 
complex environment such as the Amazon forest.

\end{abstract}

\end{frontmatter}

\section{Introduction}

Amazonia is one of the last great tropical forest domains, the largest hydrological system in the 
planet, and plays an important role in the function of regional and global climates.
Many aspects of this fragile and highly complex system remain unclear for the scientific community. 
A subject of great relevance for understanding how the Amazon terrestrial biosphere interact with the atmosphere
is the correct modeling of the turbulent exchange of heat, humidity, greenhouse gases, and other scalars 
at the interface vegetation-air. This is partly due, on one hand, to the lack of high-frequency, detailed 
{\it{in-situ}} measurements, and, on the other hand, to the fact that turbulence has been a notoriously
difficult problem to grasp. While most researchers agree that the basic physical 
aspects of the mechanically generated turbulence are described by the Navier-Stokes equations, limitations 
in computer capacity make it impossible to directly solve these equations for high Reynolds
numbers turbulent flows (fully developed turbulence), specially in a complex environment such as the
canopy of a rain forest.

Traditionally, the properties of turbulent flows are studied from the probability density
functions (PDFs) of fluctuating quantities (velocity differences $v_r(x) = v(x) - v(x+r)$,
for example), at different separation scales $r$. It is a well known characteristic of
turbulent flows that, at large scales, these PDFs are normally distributed. However, at increasingly smaller scales, 
they become strongly non-Gaussian and display tails fatter than expected for a normal process. 
This is the signature of the intermittency phenomenon: strong bursts in the kinetic energy dissipation rate. 
Since the sixties, many PDFs models have been proposed to take into account 
this feature \cite{Frisch:95,Sreenivasan:97}.  
Most of these models are based on refined versions
of Kolmogorov's original phenomenology for isotropic inertial subrange turbulence, 
such as in the lognormal \cite{Kolmogorov:62}, multifractal 
\cite{Parisi:85}, log-Poisson \cite{She:95}, 
and Levy \cite{Painter:96} models.

Recently, a new PDF model based on the non-extensive thermostatistics (NETS) formalism has been
introduced \cite{Ramos:99}. Since then, the connection between NETS and turbulence is attracting a 
growing interest \cite{Bolzan:00,Beck:00,Arimitsu:00a,Arimitsu:00b,Ramos:01a,Ramos:01b,Beck:01,Bolzan:02,Peyrard:02}. 
NETS is a generalization of classical Boltzmann-Gibbs 
thermostatistics \cite{Tsallis:88}, through the introduction of a family of non-extensive entropy functionals
$S_q$, with a single parameter $q$. These functionals reduce to the Boltzmann-Gibbs entropy as $q \rightarrow 1$.

Within the context above, the objective of this paper is twofold. First, to study the atmospheric turbulence in a complex 
environment such as the Amazon forest. In particular, we focus on the impact of the rain forest crown 
on the statistical characteristics of the atmospheric turbulence, and on how this characteristics are 
affected when one moves into the canopy, or when one passes from day/unstable to night/stable conditions.
We also investigate the connection between coherent structures and intermittency on the
statistical distribution of turbulence fluctuations. Our second goal is to test the validity of the PDF model 
based on NETS, and whether this approach can provide new insights to the study of atmospheric turbulence in
the tropics. To achieve these goals, we use fast-response experimental data obtained during the wet-season campaign of 
the Large Scale Biosphere-Atmosphere Experiment in Amazonia (known as the LBA Project), 
carried out during the months of January to March of 1999, in the southwestern 
part of the Brazilian Amazonia.

This paper is organized as follows. In Section 2 we describe the data and the experimental site. Section 3 contains
the theoretical background. Results are presented and discussed in Section 4. Finally, in Section 5 we present
our conclusions.

\section{Data and Experimental Site}

The experimental site is located in Rondonia, Brazil, roughly 3000 kilometers northwest from Rio de 
Janeiro, inside the Jaru Biological Reserve, a densely forested area with 270 thousand hectares.
Fast response wind speed measurements, in the three orthogonal directions, and temperature 
measurements were made at a sampling rate of 60 Hz, using sonic anemometers and 
thermometers. The data was gathered during an intensive micrometeorological campaign,
part of the wet-season LBA project. The experiment was carried out
during the months of January to March 1999. 
The LBA Project, acronym for Large Scale Biosphere-Atmosphere Experiment in Amazonia, is an 
international initiative led by Brazil, aimed at understanding the climatological, ecological, 
biogeochemical, hydrological functioning of Amazonia, studying the impact of land use change, 
specially deforestation, in these functions, and analyzing the interactions between Amazonia 
and the Earth system. 

The measurements were made with the help of a 66 meters micrometeorological 
tower, simultaneously at three different heights: above the canopy, at 66 m, at the canopy top, 
at 35 m, and within the canopy, at 21 m.
Two distinct measurement periods have been selected: from noon to 1:00 pm, when the forest crown 
is heated by the sun, the top of the canopy is hotter then the surroundings, and thus the above canopy 
region is unstable; and from 11:00 pm to midnight, when we have the opposite condition, and the above 
canopy region is stable. In order to verify the data quality, we applied the quality control procedure 
proposed by Vickers and Mahrt \cite{Vickers:97}. We also checked the validity of Taylor's hypothesis verifying 
the turbulence intensity inside the inertial subrange \cite{Wyngaard:77}. Finally, since we were primarily interested 
in the statistical characteristics 
of turbulence within the inertial subrange, we checked our data for the existence of a sizable scaling range. 
We also computed the approximate extension of the inertial sub-range using the value of the isotropy coefficient 
(which shall be close to one within the intertial sub-range) \cite{Kulkarni:99}.

\section{Theoretical Background}

In this paper, we adopt a generalization of the model used in our previous works \cite{Ramos:99,Bolzan:00,Ramos:01b}, 
assuming that the PDF $p_q(v_r)$ of turbulent velocity differences $v_r$ 
(and also temperature differences $T_r$) is given by \cite{Beck:01}:

\begin{equation}
\label{eq2} 
p_q(v_r) = [1 - \beta (1 - q)[|v_r|^{2 \alpha} - C
sign(v_r) (|v_r|^{\alpha} - \frac{1}{3} |v_r|^{3 \alpha}))]^{1 /
(1 - q)} / Z_q,
\end{equation}
where $C$ is a small skewness correction term, and $Z_q$ is given by 

\begin{equation}
\label{eq3} Z_q = \frac{a^{m_0 + 1}}{\alpha} B(\phi_0, \chi_0),
\end{equation}
with $B(\phi_0,\chi_0) = \Gamma(\phi_0) \Gamma(\chi_0) /
\Gamma(\phi_0 + \chi_0)$, $\phi_0 = (1 + m_0) / 2$, $\chi_0 = l -
\phi_0$, $l = \frac{1}{q - 1}$, $m_0 = \frac{1 - \alpha}{\alpha}$,
and $a = \sqrt{l / \beta}$. The parameter $\alpha$ was
chosen according to the empirical formula $\alpha = 6 - 5q$. 
The main advantage of eq. (\ref{eq2}) is to permit the use of the same PDF model for handling 
both velocity and temperature turbulent fluctuations.

Neglecting the skewness correction term, we obtain for the PDF
$n$-th moment:

\begin{equation}
\label{eq4} <|v_r|^n> = a^{m_n - m_0}
\frac{B(\phi_n,\chi_n)}{B(\phi_0, \chi_0)},
\end{equation}
where $\phi_n = (1 + m_n) / 2$, $\chi_n = l - \phi_n$ and $m_n =
\frac{(n + 1) - \alpha}{\alpha}$. 

The parameters $q$ and $\beta$ determine the shape of the PDF and
are obtained through eq. (\ref{eq4}), measuring the values of two moments (or related quantities) at each scale 
(for example, the variance, $<|v_r|^2>$ and the kurtosis, $K_r = \frac{<|v_r|^4>}{<|v_r|^2>^2}$). 

The parameter $q$ depends on $K_r$ through the equation: 

\begin{equation}
\label{eq5} K_r = \frac{B(\phi_4,\chi_4)
B(\phi_0,\chi_0)}{B(\phi_2, \chi_2)^2},
\end{equation}

Particularly for $<|v_r|^2> = 1$, $\beta$ is given by 

\begin{equation}
\label{eq6} \beta = l
[\frac{B(\phi_2,\chi_2)}{B(\phi_0,\chi_0)}]^{2 / (m_2 - m_0)} ,
\end{equation}

We remark that the kurtosis depends only on the entropic parameter. It is well known that large values of $K_r$ 
are a signature of intermittency \cite{Sreenivasan:97}. 
Thus, $q$ can be used as a measure of intermittency in turbulent flows \cite{Bolzan:02}. 

We also note that if we assume a scaling of the moments $<|v_r|^n>$ of $v_r$ as $r^{\varsigma_n}$
(which is valid for inertial subrange scales, for sufficiently
high Reynolds number), the scale variation of $q$ and $\beta$ can
be {\it{computed}} (rather than measured). For this, we shall use the scaling relation together with
equations (\ref{eq4}) and (\ref{eq5}) to extrapolate the experimental values of $q$ and $\beta$, at a given reference
scale (say, the Kolmogorov scale, $\eta$). This
extrapolation procedure can be extended over a much wider range
of scales \cite{Bolzan:02} with respect to the inertial subrange, by using the concept
of extended self-similarity proposed in \cite{Benzi:95}. This
approach requires to numerically solve the Kolmogorov equation
using, as initial condition, the observed value of $q$ and $\beta$, at a reference scale. 

\section{Results and Discussions}

\subsection{Wind Velocity Data}

\begin{figure}
\begin{center}
\includegraphics[width=12cm]{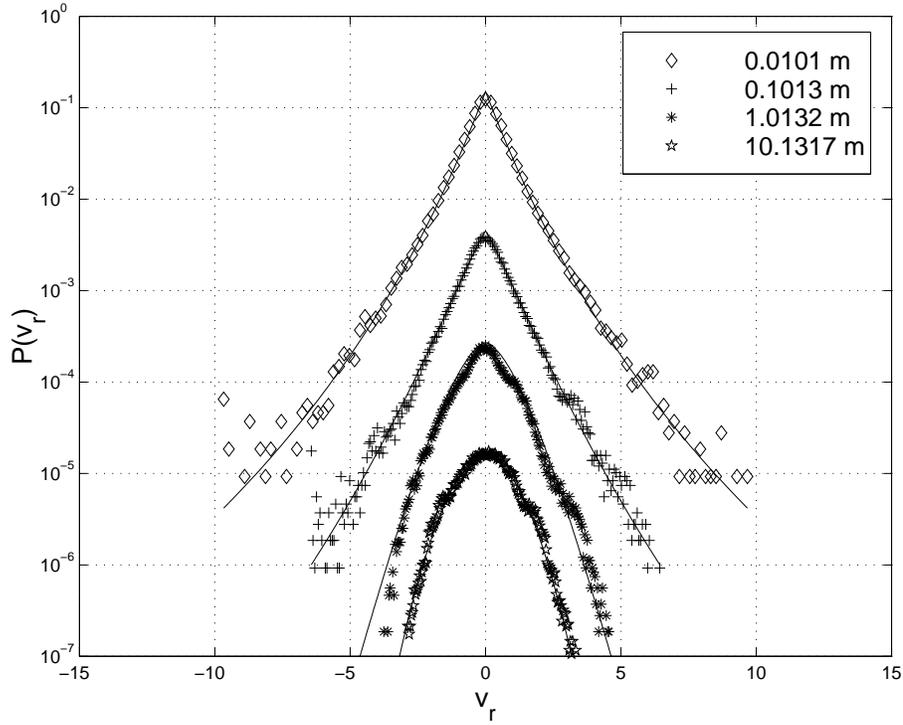}
\caption{Standardized experimental and theoretical (solid lines) probability distributions of vertical
velocity differences at the four spatial scales, for daytime above canopy data.}
\label{fig2}
\end{center}
\end{figure}

\begin{figure}
\begin{center}
\includegraphics[width=12cm]{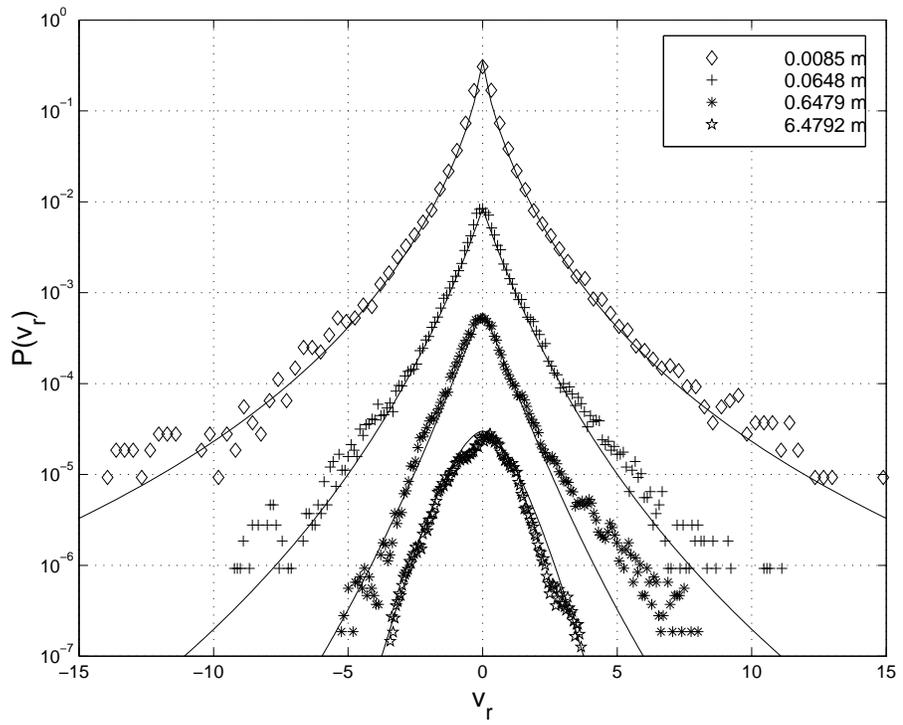}
\caption{The same as in Figure 1, but for within canopy data.}
\label{fig3}
\end{center}
\end{figure}

Figures \ref{fig2} and \ref{fig3} present semilogarithmic plots of probability distributions of 
daytime normalized vertical velocity differences $v_r = w(x)-w(x+r)$, at four different scales,
properly rescaled and vertically shifted for better visualization. 
Figure \ref{fig2} data was measured above the canopy, inside the transition sublayer. Figure
\ref{fig3} data was measured approximately 14 meters below the canopy top. 
Overall, we observe that the theoretical results (solid lines) are in good agreement with measurements across
spatial scales spanning three orders of magnitude and a range of up to 10 standard deviations, 
including the rare fluctuations at the tail of the distributions. The transition from large-scale
Gaussian behavior to a power-law form as $r$ decreases is quite evident and well reproduced by the model. 
At small scales, the distributions have tails larger than that expected for a normal process and a spiky shape 
near the origin, an indicative of intermittency. We obtained
similar agreement for the PDFs of longitudinal velocity $u$ differences (not shown in the text). 

\begin{figure}
\begin{center}
\includegraphics[width=14cm]{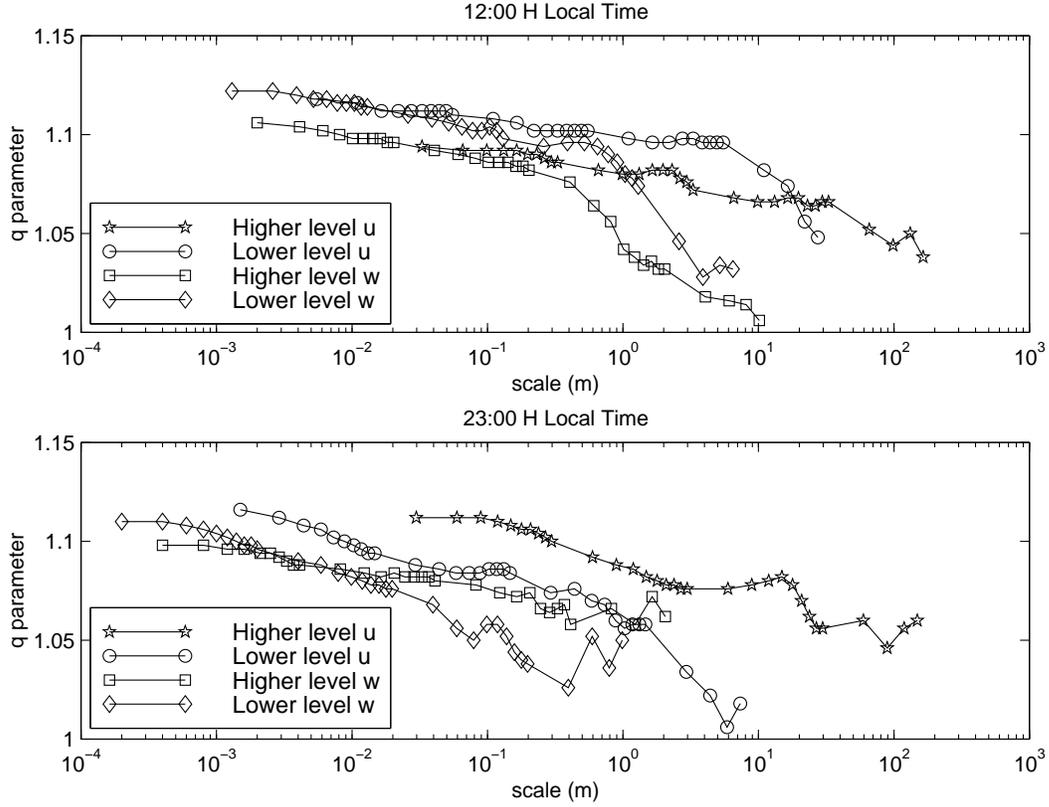}
\caption{Scale variation of the parameter $q$ for vertical velocity
($w$) and longitudinal velocity ($u$) measured above and within the 
canopy, under diurnal conditions (top) and nocturnal conditions (bottom).}
\label{fig4}
\end{center}
\end{figure}

Comparing the histograms of Figures \ref{fig2} and \ref{fig3}, we note that the kurtosis is consistently 
higher within the canopy under diurnal conditions, regardless the scale. To investigate with more detail 
this behavior, we plot in Figure \ref{fig4}, the scale variation of the entropic 
parameter for $u$ and $w$ data, above and within the canopy, under diurnal and
nocturnal conditions. A few points can be highlighted from these results.

First, we remark that all curves display a similar pattern: from a saturation value, the entropic 
parameters $q_u$ and $q_w$ decrease as $r$ grows. Theoretically, $q$ should 
tend to 1 at the integral scale and beyond. A similar trend has also been observed in a Couette-Taylor flow 
experiment \cite{Beck:01}.

Second, we note that the statistical characteristics of $u$ and $w$ wind-velocity components are not 
the same, mainly at larger scales. This result was somehow expected considering that our data was 
measured in a "noisy" real atmosphere, close to a very complex surface, such as the Amazon rain 
forest canopy.

Third, we observe that, indeed, under diurnal conditions, the entropic parameter is 
consistently higher within the canopy. However, this bias towards higher levels
of intermittency found in low level data disappears under nocturnal conditions, as shown in Figure \ref{fig4}b. 
In order to explain this behavior, it is essential to examine the cyclic variation in the thermal stability 
regimes above and within the canopy along a typical day, and the role of the forest canopy in this 
process. 

Schematically, during the day, dense forest canopies store heat in their highest parts due to incoming solar 
radiation flux. Hence, under daytime conditions, the above canopy region is hotter than the 
surroundings, and, thus, unstable. On the other hand, the region within the canopy is stable. There is a 
downward flux of turbulent kinetic energy (TKE), which is mostly produced by mechanical shear of the flow 
next the canopy. 
During the night, the energy budget is dominated by long-wave infrared radiation. 
Thus, the forest crown looses heat, the stability profile is reversed and stable conditions predominate above the canopy, 
and lightly unstable conditions may occur within the canopy. Next to the ground, there is a small upward 
flux of TKE generated by thermally induced local flows. This cyclic process determines variations on the thermodynamic 
structure of the canopy, which influence the turbulent transfer processes in this environment. 
This analysis suggests a simple scenario to explain the different intermittency regimes observed in the data.
In this scenario, the forest crown act as a filter, breaking down large vortices while allowing smaller 
ones to pass through the canopy. This filtering process also explains why stable regions have a higher
intermittency level than unstable ones. We remark that such eddy-filter character of forested canopies 
has already been observed by other authors
\cite{Shuttleworth:85,Fitzjarrald:90b,Kruijt:00}. 

\begin{figure}
\begin{center}
\includegraphics[width=12cm]{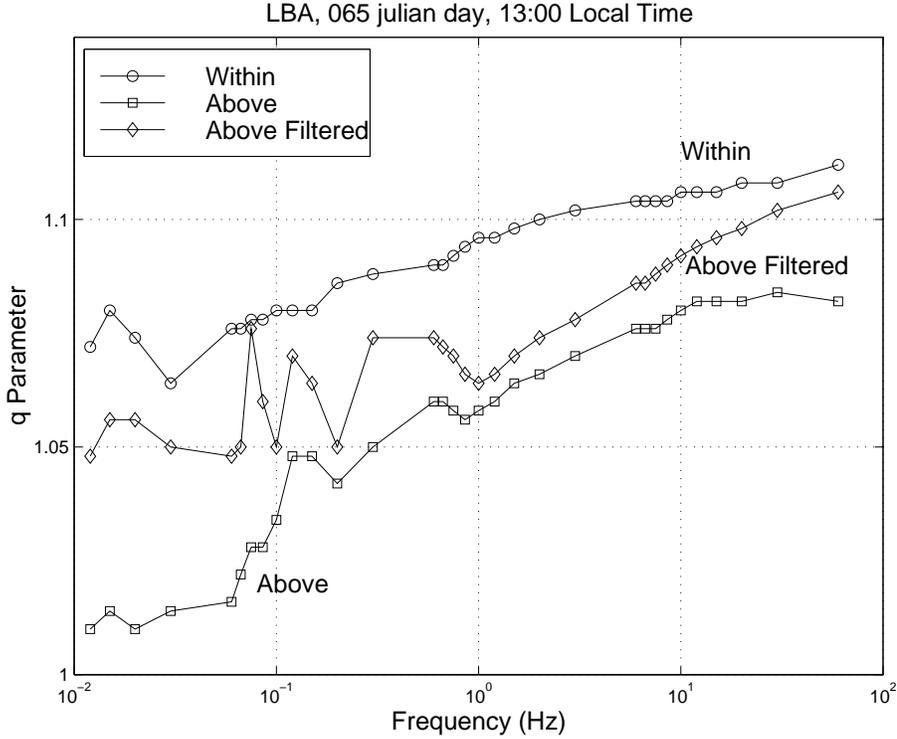}
\caption{Entropic parameter corresponding to daytime above canopy, high-pass filtered above canopy, and 
within canopy data. }
\label{nova1}
\end{center}
\end{figure}

In order to test this scenario, we high-pass filtered the original daytime above-canopy signal, and measure 
at each scale, the corresponding entropic parameter. As we can see Figure \ref{nova1}, this procedure increases 
the signal kurtosis, resulting in PDFs that are more similar to those found within the canopy during the day. 
This result provides evidence that indeed the forest crown has a filtering effect on large eddies, what impacts
the intermittency level of the remaining velocity field. 
Naturally, the real scenario is much more complex than that, and is difficult 
to establish a general and simple pattern for all turbulent fluctuations in the actual atmosphere. 
For example, the momentum exchange process between the atmospheric flow above and within the canopy is not continuous 
in time but characterized by strong intermittent transfers, associated with the action of the so-called coherent structures, 
and characterized by a sweep and an ejection phases \cite{Gao:89}.

\subsection{Temperature Data}

\begin{figure}
\begin{center}
\includegraphics[width=12cm]{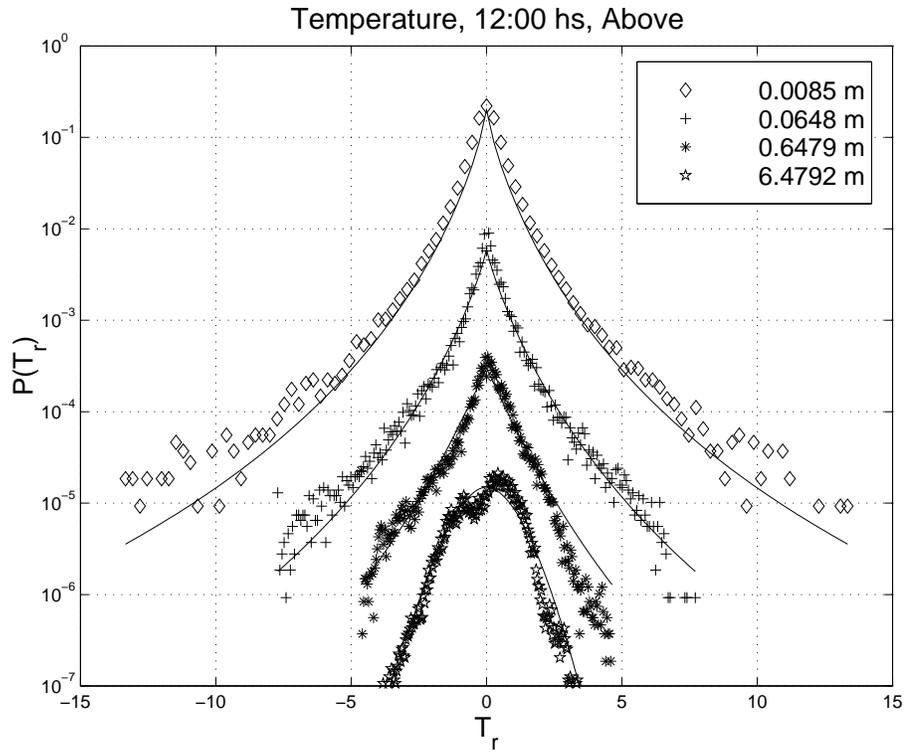}
\caption{Standardized experimental and theoretical (solid lines) PDFs of temperature 
differences at four spatial scales, for diurnal above canopy data.}
\label{nova2}
\end{center}
\end{figure}

\begin{figure}
\begin{center}
\includegraphics[width=12cm]{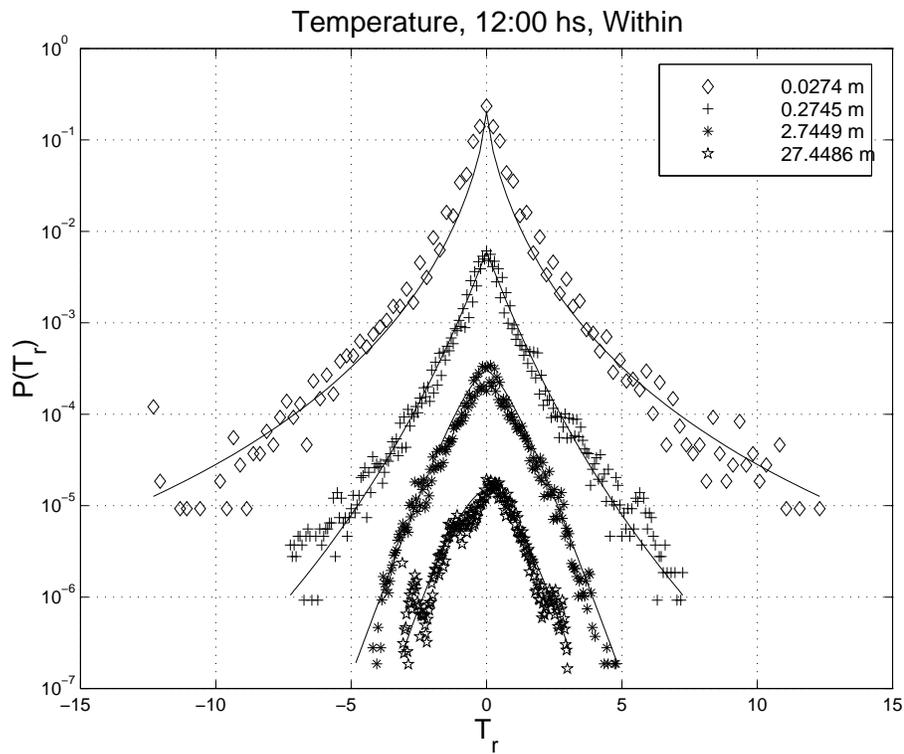}
\caption{The same as in Figure 5, but for within canopy data.}
\label{nova3}
\end{center}
\end{figure}

We also tested the PDF model given by eq. (\ref{eq2}) with temperature data.
To illustrate, in Figures \ref{nova2} and \ref{nova3} we compare the theoretical PDFs with the experimental 
histograms of daytime normalized temperature differences, above and within the canopy, at four different scales. 
Again, for each scale, we estimated the variance and the kurtosis, and then 
computed the parameters $q$ and $\beta$. Overall, we observe that the theoretical results (solid lines) are in good agreement with 
measurements, mainly at the smaller spatial scales. The transition from a power-law form at smaller 
scales to large-scale Gaussian behavior, as the scale increases, is less evident than in the velocity 
histograms but is also present. Comparing both histograms, we also note that, at smaller scales, they are quite similar, 
but at larger scales, the above canopy data appears to be more spiky, with heavier tails.

\begin{figure}
\begin{center}
\includegraphics[width=14cm]{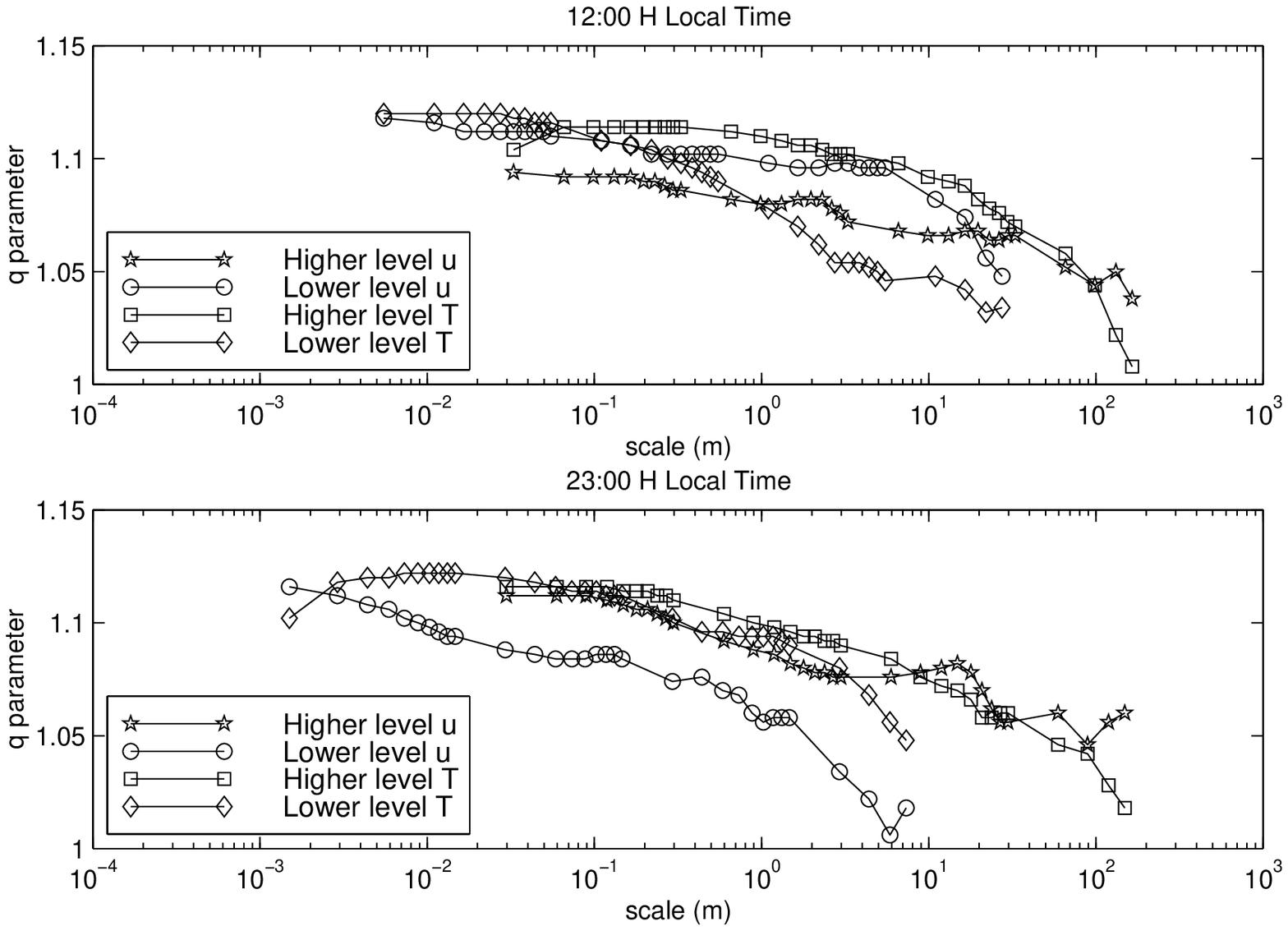}
\caption{Scale variation of the parameter $q$ for longitudinal
velocity ($u$) and temperature (T) measured above and within the canopy,
under diurnal conditions (top) and nocturnal conditions (bottom).}
\label{fig5}
\end{center}
\end{figure}

As we did previously, we study this trend plotting in Figure \ref{fig5}, the scale variation of the entropic parameter for 
temperature data, above and within the canopy, under nocturnal and diurnal conditions.
Again, we observe that velocity and temperature curves display a similar pattern: from a saturation value, $q$  
decreases as one goes to larger scales. However, we also remark that, under both diurnal and nocturnal conditions, the entropic parameter 
is higher above the canopy. In other words, the temperature signal appears to be more intermittent above the forest crown, 
regardless the stability conditions. In this case, it is worthwhile to ask why this behavior is different from that 
observed in the velocity data, which display different patterns for day and night. 

One possible answer is the existence of large-scale, ramp-like, coherent structures in the 
temperature fields. These structures are responsible for most of the sensible 
heat transport through the canopy \cite{Chen:97}. Since the strongest shear and thermal 
gradients are located above the forest crown, we expect that ramp-like structures will be more apparent 
above vegetated canopies than within them \cite{Paw:92}.
These large scale, coherent structures are influenced by the local boundary conditions and, through 
their interaction with smaller scales, affect inertial sub-range properties measured by $q$. 
If correct, this scenario suggest that a "universal model" of turbulence intermittency is very difficult 
to be defined or may even not exist for canopy flows. These findings have
implications in the development of subgrid model of Large Eddy Simulation (LES), now widely used to assess $CO_2$ 
exchange \cite{Albertson:01}, 
which are primarily based on Kolmogorov type cascades or simplistic energy backscatter corrections \cite{Meneveau:00}, 
as they are not capable to capture the effect of large-scale motion on canopy sublayer inertial range. 

\begin{figure}
\begin{center}
\includegraphics[width=12cm]{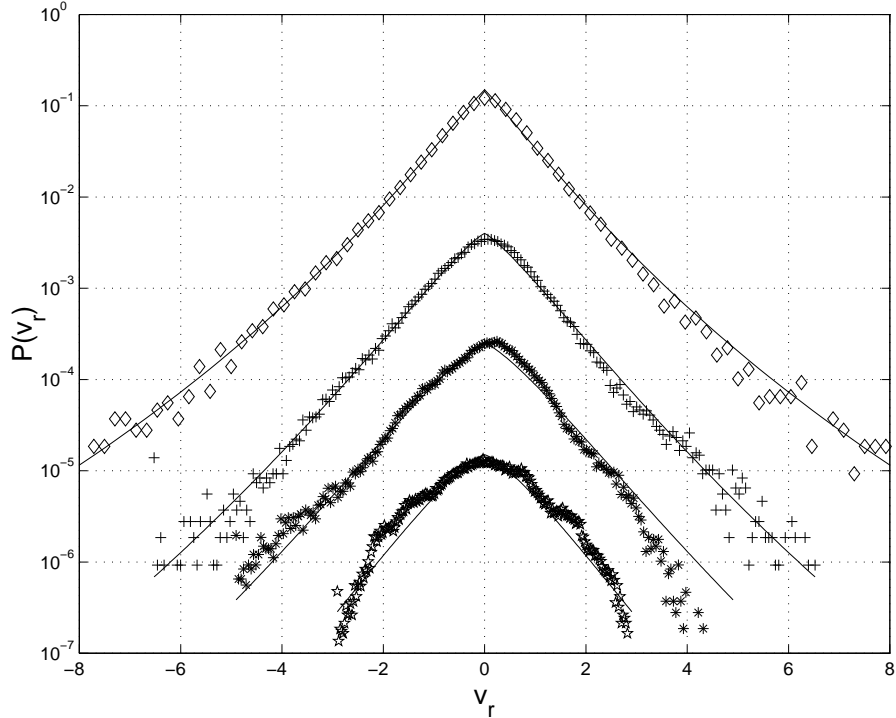}
\caption{Histograms of the original turbulent wind-velocity data.}
\label{nova5}
\end{center}
\end{figure}

\begin{figure}
\begin{center}
\includegraphics[width=12cm]{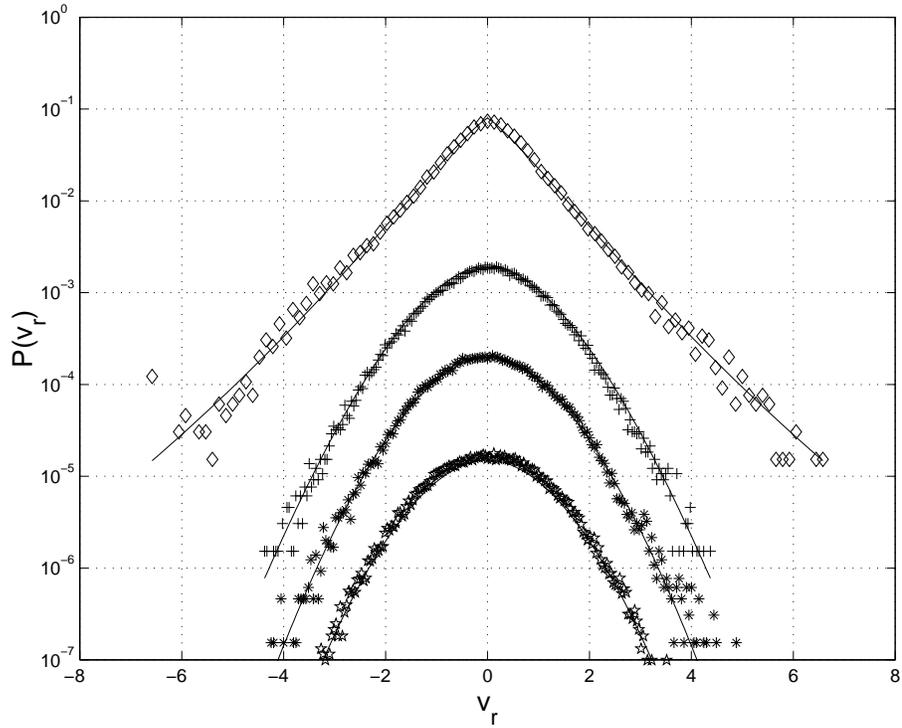}
\caption{Histograms of the filtered decorrelated wind-velocity data.}
\label{nova6}
\end{center}
\end{figure}

The impact of such large scale, coherent structures can be well illustrated decomposing the 
turbulent data, by means of Haar wavelet filtering \cite{Katul:94b}, in a coherent, intermittent 
part, and an incoherent, 
structureless part. This is performed by removing  from the original signal all 
wavelet coefficients that are larger than a given threshold. Comparing the experimental histograms of the original 
signal (Figure \ref{nova5}) and the remaining incoherent signal (Figure \ref{nova6}), we observe that the incoherent, 
decorrelated velocities display PDFs that are roughly Gaussians, where the effect of intermittency has almost 
disappeared. Naturally, the corresponding power spectrum (not shown here) is much more similar to that of a 
white noise.

\begin{figure}
\begin{center}
\includegraphics[width=12cm]{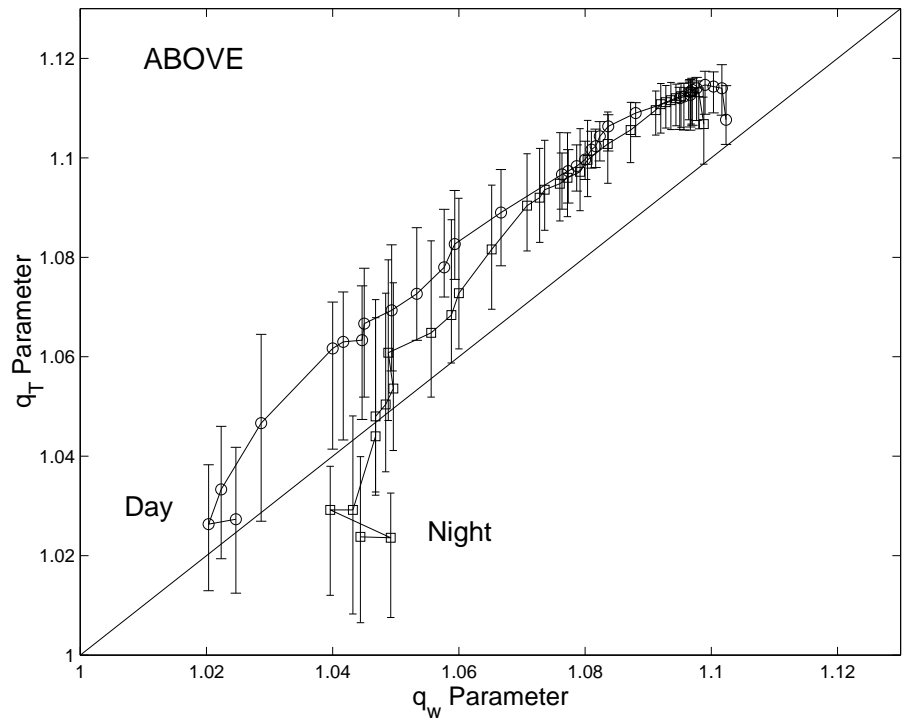}
\caption{Pairs of $q_w$ and $q_T$, simultaneously measured at different scales, 
for daytime and nighttime conditions, above the forest canopy.}
\label{nova7}
\end{center}
\end{figure}

\begin{figure}
\begin{center}
\includegraphics[width=12cm]{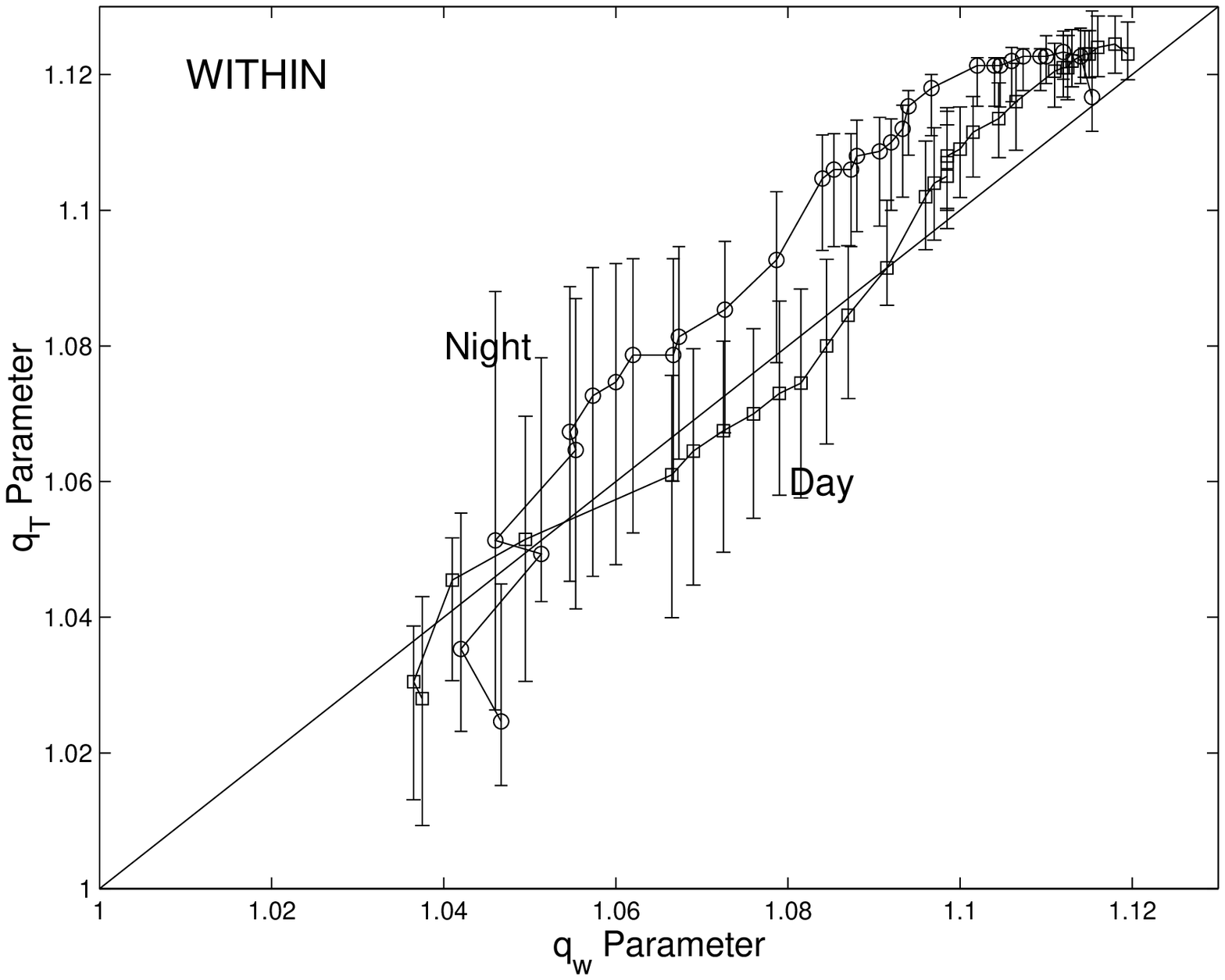}
\caption{Pairs of $q_w$ and $q_T$, simultaneously measured at different scales, 
for daytime and nighttime conditions, within the forest canopy.}
\label{nova8}
\end{center}
\end{figure}

Although the values of $q$ for velocity and temperature are highly correlated, they also convey 
information about different aspects of the turbulent flow: $q_w$  about the momentum exchange 
process through the canopy, and $q_T$  about the transport of sensible heat.
Thus, properly combining the information on the two entropic parameters, make it possible to assess the stability 
conditions of the atmosphere. In Figures \ref{nova7} and \ref{nova8}, we plot pairs of 
$q_w$ and $q_T$, simultaneously measured at different scales, for daytime and nighttime conditions, above and 
within the forest canopy. As we can see, two different stability regimes are clearly visible.

Above the canopy (Figure \ref{nova7}), the atmosphere is thermally unstable during the day ($q_w < q_T$), and almost 
stable at large scales during the night ($q_w \approx q_T$), although there are evidence of instabilities at smaller scales. 
On the other hand, within the forest (Figure \ref{nova8}), the atmosphere is slightly unstable during 
the night ($q_w < q_T$) and stable during the day ($q_w \approx q_T$). In contrast with the usual stability criteria 
\cite{Lumley:64,Mahrt:89}, 
which are essentially global, the main advantage of the present approach is to assess the local atmospheric conditions, 
at different range of scales.

\section{Conclusions}

In this paper, we discussed the impact of a rain forest canopy on the statistical characteristics of 
atmospheric turbulence. This issue is of particular interest for understanding on how the Amazon 
terrestrial biosphere interact with the atmosphere. For this, we used a probability density function (PDF) model of 
velocity and temperature differences based on Tsallis' non-extensive thermostatistics (NETS). This new
approach allow us to use a single PDF model to describe both the turbulent velocity and temperature 
differences in the turbulent flow.

We compared theoretical results with experimental data measured in a 66 m micrometeorological 
tower, during the wet-season campaign of the Large Scale Biosphere-Atmosphere Experiment in 
Amazonia (LBA). We investigated in detail how the value of Tsallis' entropic parameter is affected 
when one moves into the canopy, or when one passes from day/unstable to night/stable conditions.
We observed that the forest crown break down large vortices, allowing only smaller eddies to pass through the 
canopy, what increases the intermittency level of the remaining turbulent velocity data. We also found that
large-scale, ramp-like, coherent structures in the temperature fields affect inertial sub-range 
properties increasing the kurtosis of the temperature signal above the canopy. Finally, we showed that combining 
the information on $q_w$ and $q_T$ make it possible to assess the stability conditions of the atmosphere within 
and above the canopy.

In conclusion, we might say that the new approach described in this paper, based on NETS, provides interesting 
insights on different aspects of the atmospheric turbulence in an complex environment 
such as an Amazon rain forest. In this context, Tsallis' entropic parameter emerges as a measurable quantity 
which can be used to objectively quantify intermittency buildup in turbulent atmospheric flows.

\section{Acknowledgments}

This work is part of The Large Scale Biosphere-Atmosphere
Experiment in Amazonia (LBA) and was supported by the Funda\c
c\~ao do Amparo a Pesquisa do Estado de S\~ao Paulo
(FAPESP)/Brazil-process 1997/9926-9. This work has also
been supported by CNPq and CAPES - Brazil.

\bibliographystyle{unsrt}

\bibliography{refe}

\end{document}